# Accelerating and Stopping Resistance Drift in Phase Change Memory Cells via High Electric Field Stress

R. S. Khan,[1] A. H. Talukder,[1] F. Dirisaglik[1,2], H. Silva,[1] and A. Gokirmak[1]
[1]*Department of Electrical and Computer Engineering, University of Connecticut, Storrs, Connecticut 06269, USA*
[2]*Department of Electrical and Electronics Engineering, Eskisehir Osmangazi University, Eskisehir 26480, Turkey*

We observed resistance drift in 125 K – 300 K temperature range in melt quenched amorphous $Ge_2Sb_2Te_5$ line-cells with length x width x thickness = ~500 nm x ~100 nm x ~ 50 nm. Drift coefficients measured using small voltage sweeps appear to decrease from 0.12 ± 0.029 at 300 K to 0.075 ± 0.006 at 125 K. The current-voltage characteristics of the amorphized cells measured in the 85 K – 300 K using high-voltage sweeps (0 to ~25 V) show a combination of a linear, low-field exponential and high-field exponential conduction mechanisms, all of which are strong functions of temperature. The very first high-voltage sweep after amorphization (with electric fields up to ~70% of the breakdown field) shows clear hysteresis in the current-voltage characteristics due to accelerated drift, while the consecutive sweeps show stable characteristics. Stabilization was achieved with 50 nA compliance current (current densities ~$10^4$ A/cm$^2$), preventing appreciable self-heating in the cells. The observed acceleration and stoppage of the resistance drift with the application of high electric fields is attributed to changes in the electrostatic potential profile within amorphous $Ge_2Sb_2Te_5$ due to trapped charges, reducing tunneling current. Stable current-voltage characteristics are used to extract carrier activation energies for the conduction mechanisms in 85 K – 300 K temperature range. The carrier activation energy associated with linear current-voltage response is extracted to be 331 ± 5 meV in 200 - 300 K range, while carrier activation energies of 233 ± 2 meV and 109 ± 5 meV are extracted in 85 K to 300 K range for the mechanisms that give exponential current-voltage responses.

Phase change memory (PCM) has recently entered the consumer memory market as a 3D-integrated high-density 2-terminal resistive non-volatile memory technology[1]. The phase change materials, such as $Ge_2Sb_2Te_5$ (GST), used to form the active regions of PCM cells are rapidly switched between the highly-conductive crystalline state and highly-resistive amorphous state using short (ns) electrical pulses[2,3]. PCM cells experience high temperatures, high thermal gradients, high current densities, and high electric fields in small dimensions,[4,5] in short timescales, while the crystals nucleate, grow[6,7], and amorphize during the reset and set operations[8]. Even though the cell operation is rather complicated compared to conventional electronic devices and the behavior of materials is not fully understood yet, PCM is demonstrated to be a reliable and cost-effective non-volatile memory technology with cell endurances that go beyond $10^{12}$ cycles[9]. PCM can be integrated with CMOS at the back-end-of-the line.

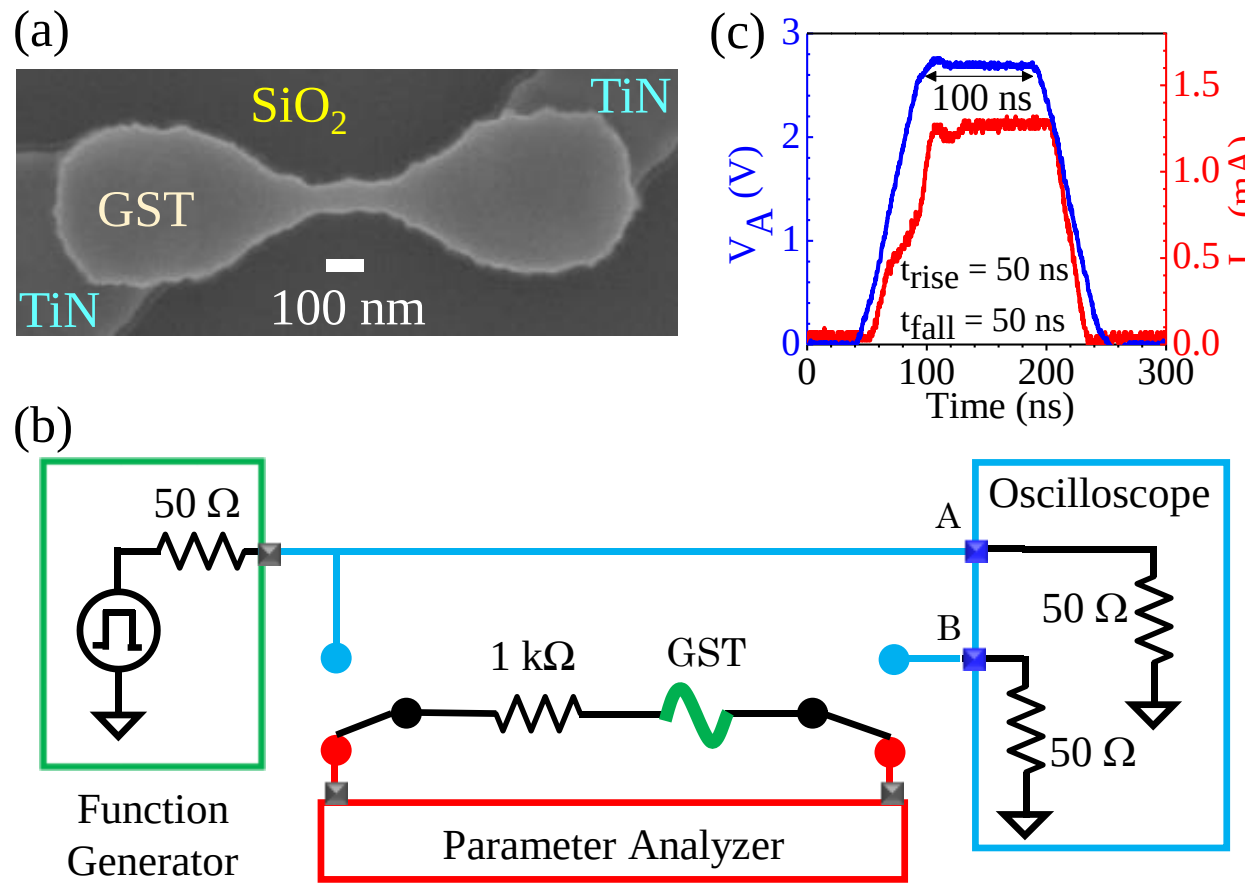

Fig. 1: Scanning electron microscopy (SEM) image of a GST line-cell with TiN bottom contacts (a). Electrical measurement setup (b). The applied reset voltage pulse and the measured current through the device during reset pulse (c).

One important scientific question that remains open regarding PCM is the physical phenomena that give rise to upward resistance drift in amorphous phase change materials, which is the spontaneous increase in resistance with time following a power-law behavior[10]:

$$R = R_0 \left(\frac{t}{t_0}\right)^{\upsilon}, \quad (1)$$

where $R$ and $R_0$ are the cell resistances at time $t$ and $t_0$ (immediately after amorphization) and $\upsilon$ is the drift coefficient. Additionally, the field and temperature dependent electrical conductivity of metastable amorphous phase of phase change materials are not fully characterized yet[11].

In this work, we use two-terminal GST line-cells (Fig. 1a) to perform current-voltage (I-V) measurements to understand the physical phenomena that give rise to resistance drift and electric field and temperature dependence of electrical conductivity of amorphous (a-) GST in 85 K to 300 K range. The line-cells used for these experiments are length ($L$) x width ($W$) x thickness ($t$) = ~400-500 nm x ~120-150 nm x ~ 50 nm with 250 nm thick bottom metal (Tungsten with Ti/TiN liner) contacts on 700 nm thermally grown $SiO_2$. The line-cells were formed by patterning a 50 nm GST thin film deposited using sputtering, optical lithography and reactive ion etching (RIE). GST is capped with 10 nm $Si_3N_4$ for passivation (Fig. 1a)[12]. The cells were annealed to transition from the as deposited a-GST to the low-resistivity hexagonal closed packed (HCP) state. Electrical measurements are performed on the cells after amorphizing each cell using a single 100 ns pulse with 50 ns rise and fall times inside a Janis

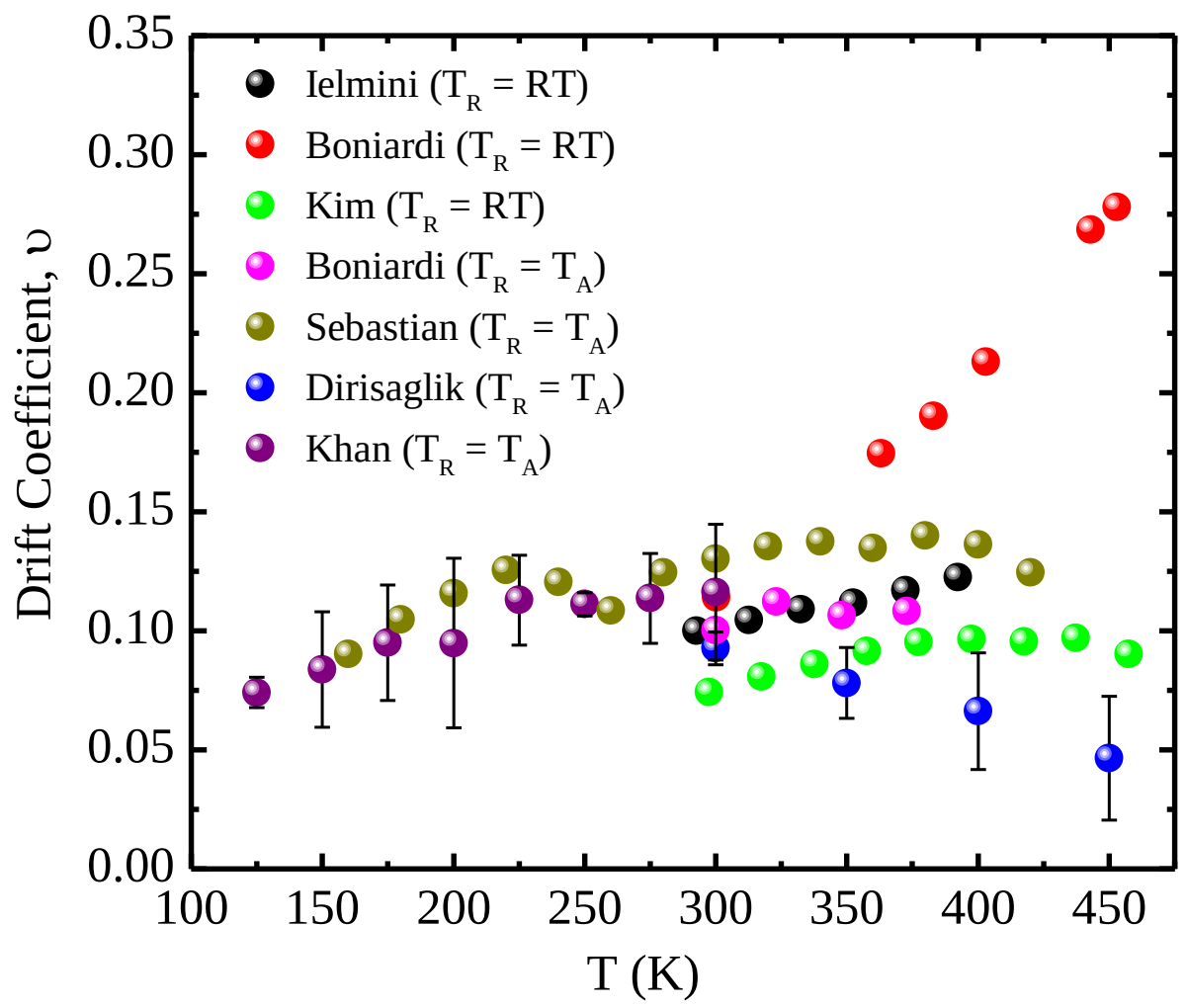

Fig. 2: Temperature-dependent behavior of drift coefficient of amorphous phase change materials[13–18]. $T_R$ and $T_A$ refers to read temperature and anneal temperature respectively.

Cryogenics probe station under vacuum (Fig. 1b,c). A 1 kΩ load resistor mounted on a probe arm was used to limit the current during amorphization. The pulse duration and rise/fall times are chosen to ensure amorphization of the cells without having distortions in the waveforms due to reflections and parasitic contributions in the electrical setup. After termination of the pulse, the relay was manually disconnected (time elapsed in disconnecting the relay is ~100 s) and DC current-voltage (I-V) sweeps were performed using an Agilent 4156C parameter analyzer (Fig. 1b) using small voltages (-0.3 V to 0.3 V) to characterize drift as done in other works[13–18], and using large voltages (0 – 25 V) with compliance currents ($I_{comp}$) in the order of 50 nA to characterize the high-field behavior of the cells. We used the current measured at the ground terminal for the analysis, which is less susceptible to leakage currents in the experimental setup. Measurements are performed in dark.

Our small-voltage (-0.3 V to 0.3 V) measurements yielded linear response with measurable current levels in 125 K to 300 K temperature range which allow characterization of resistance drift up to $10^4$ s.[13] The drift coefficients we calculate in this range show approximately 40% decrease as the temperature is lowered from 300 K ($\upsilon = 0.12 \pm 0.029$) to 125 K ($\upsilon = 0.075 \pm 0.006$), consistent with a previous observation reported in the literature[16] (Fig. 2). One of the current explanations for resistance drift is thermally activated structural relaxation, a local low-activation viscoelastic deformation of the bond network that changes the bond structure and arrangement of the atoms[19,20]. At low temperatures, the structural relaxation is expected to be significantly slower as the kinetic energy in the system is significantly lower. Carrier activation is also an exponential function of temperature in semiconductors, making the phase change materials significantly more resistive at lower temperatures. Structural relaxation is expected to give rise to a change in the trap densities and hence change the band structure. However, the electrostatic variations introduced by the filling and the emptying of these traps is not considered in this explanation. High-field experiments can provide additional information about charging and discharging of the traps and their contribution on resistance drift.

We observe a hysteresis loop in the very first high-field measurement for each device (Fig. 3) where the voltage is continuously swept from 0 V up to 25 V with 0.1 V steps and back to 0 V, with a current compliance set to 50 nA. In the 85 K to 200 K range, the consecutive measurements do not show appreciable hysteresis and the I-V characteristics stabilize within 5 sweeps (Fig. 3) to a well-defined combination of a linear, a low-field exponential and a high-field exponential response:

$$I = \frac{1}{R_{Lin}}V + I_0(e^{\alpha V} - 1) + I_1 e^{\beta V}, \quad (2)$$

where $R_{Lin}$ represents the linear resistance, $I_0$ and $\alpha$ define the low-field exponential and $I_1$ and $\beta$ define the high-field exponential behavior (Fig. 4). The upswing of the very first measurement shows a deviation from this behavior starting at ~14 V @ 85K, indicating an accelerated change in the cells starting at electric fields in the order of 30 MV/m and currents < $10^{-10}$ A. Since these current levels are very low, the cells are not expected to experience accelerated structural relaxation due to self-heating in the whole device. The stabilization of the I-V characteristics points to dramatic suppression or complete stoppage of resistance drift.

We performed sets of high-field experiments in the ~85 K to 300 K range, amorphizing the cells at the lowest temperature and performing 5 consecutive high-voltage sweeps at each temperature point (Fig. 4) to investigate stability of the devices as the temperature is increased, as well as the electric-field and temperature dependence of electrical conductivity. While the first measurement at each temperature point showed slight hysteresis, consecutive measurements were very stable and the 5th measurement at each temperature point was used for the analysis. We fit the data in the low-field range (as indicated in Fig. 5) to a linear and low-field exponential, including an offset term ($I_{offset}$) to account for the current leakage and charging in the setup, which are typically < $10^{-13}$ A in the highest current sensitivity range:

$$I_{lowE} = \frac{1}{R_{Lin}} V + I_0 (e^{\alpha V} - 1) + I_{offset}, \quad (3)$$

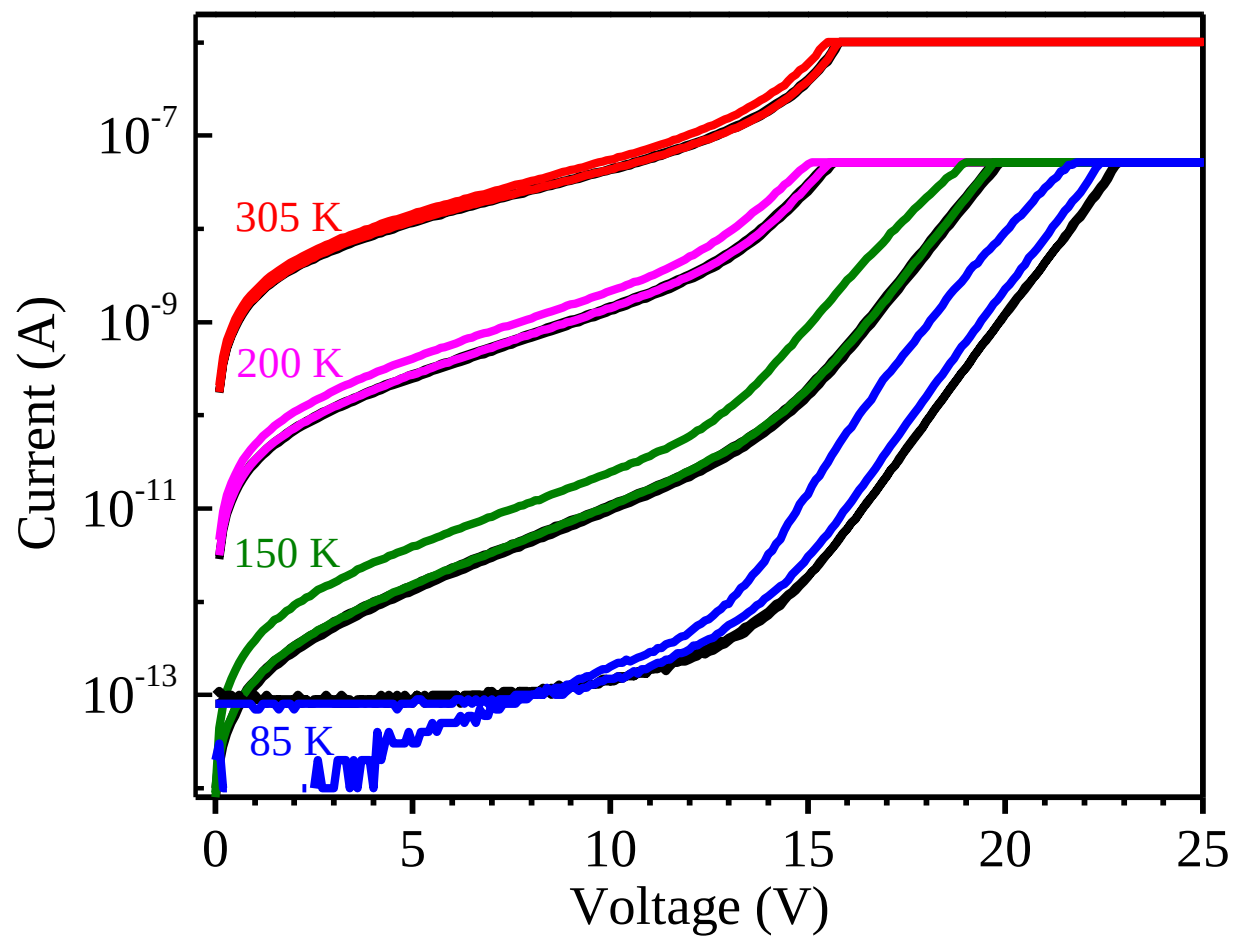

Fig. 3: First (85 K: blue, 150 K: green, 200 K: magenta, and 305 K: red) and fifth (black) current-voltage sweeps after amorphization. For 305 K, 1 μA compliance is used. Rest of the temperatures use 50 nA compliance.

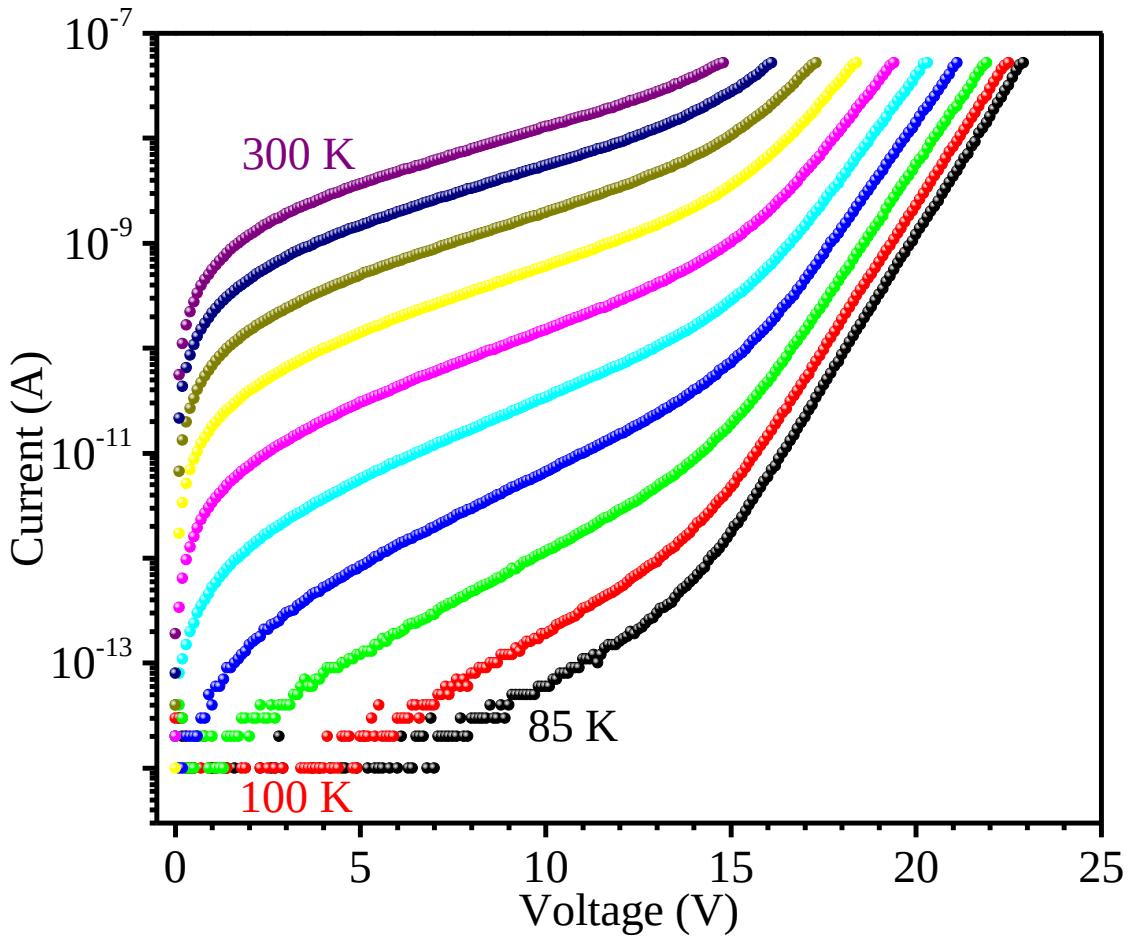

Fig. 4: Temperature dependent current vs. voltage (I-V) characteristics with 50 nA compliance current after amorphization and stabilization (5 sweeps) of a *W* x *L* ~ 150 nm x 490 nm line cell at 85 K. The data shown here are average of up-swing and down-swing of the last sweep at each temperature point.

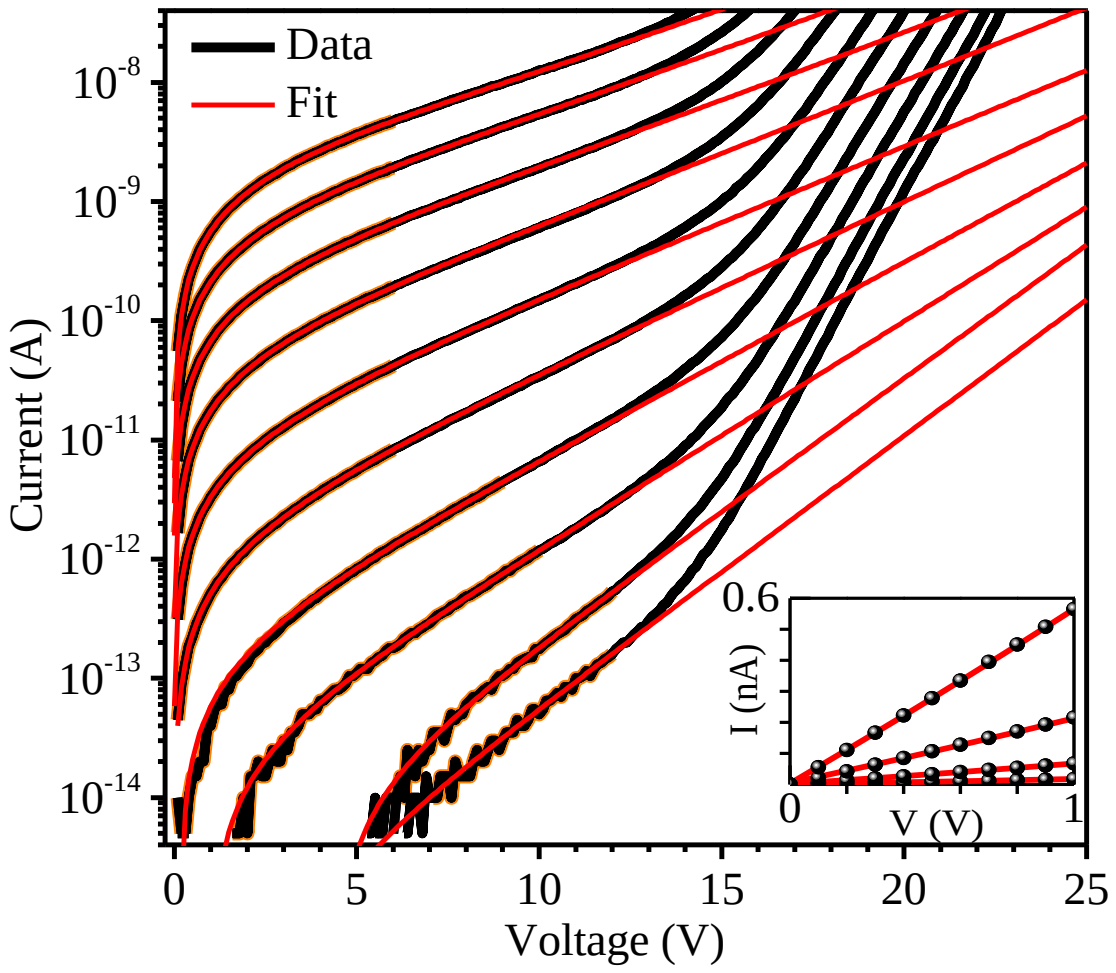

Fig. 5: Fitting of low field current extrapolated to the full range. The orange region highlights the part of the data used for fitting. The inset shows the data and fit in the 0 to 1V range in linear scale.

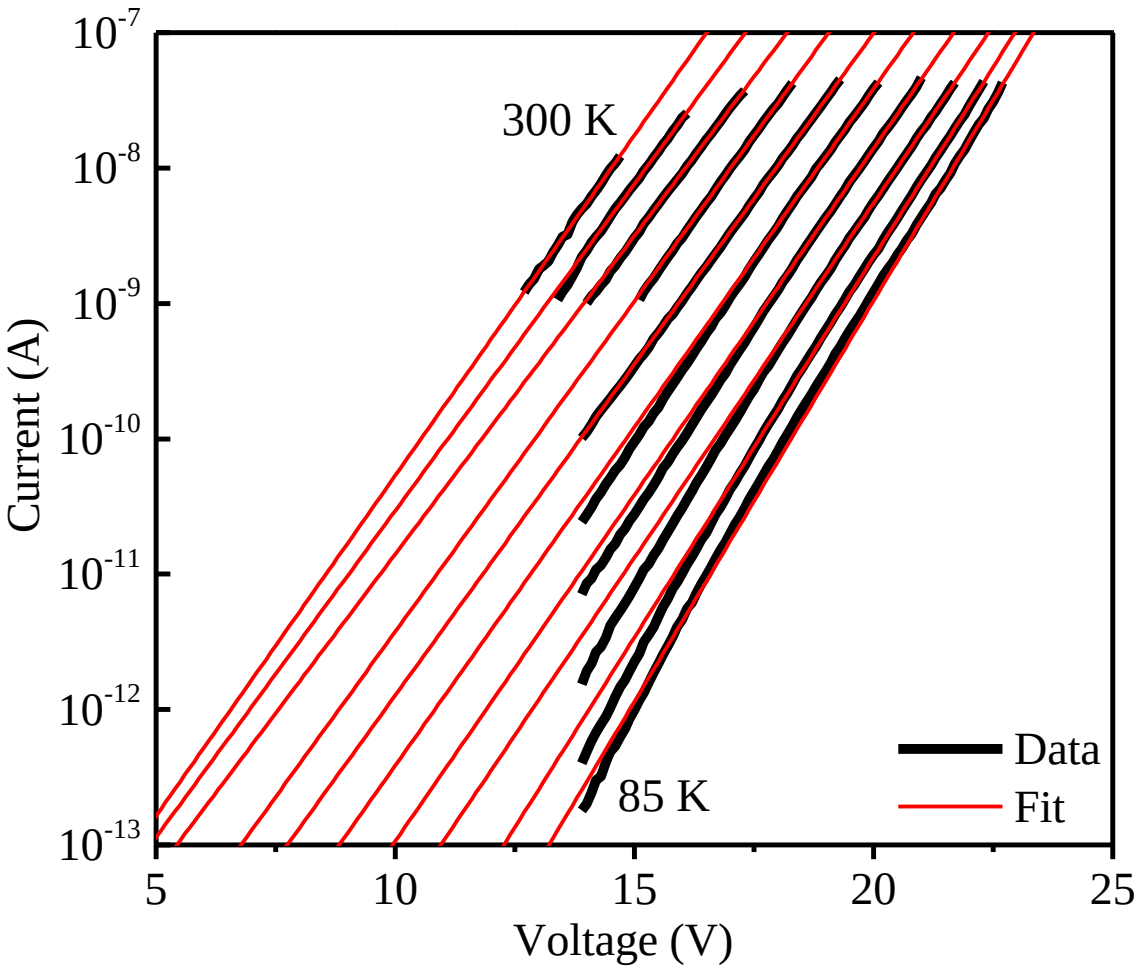

Fig. 6: High field data (raw data- low field fit, black) and fits (red).

The function obtained from the fit, extrapolated to the full voltage range, is subtracted from the experimental data to generate the high-field responses, which are then fit to simple exponential functions (Fig. 6).

The linear and the low-field exponential terms show very strong temperature dependence and tend to dominate for I < 50 nA at higher temperatures (T ≥ 225 K) (Fig 7). Since the $I_{comp}$ is reached at relatively low voltages, the devices do not experience very high-fields for higher temperatures.

The temperature dependence of the linear coefficient ($1/R_{Lin}$) can be described using an Arrhenius relationship:

$$\frac{1}{R_{Lin}} = \frac{1}{R_0} e^{-\frac{E_{aL}}{kT}}, \quad (4)$$

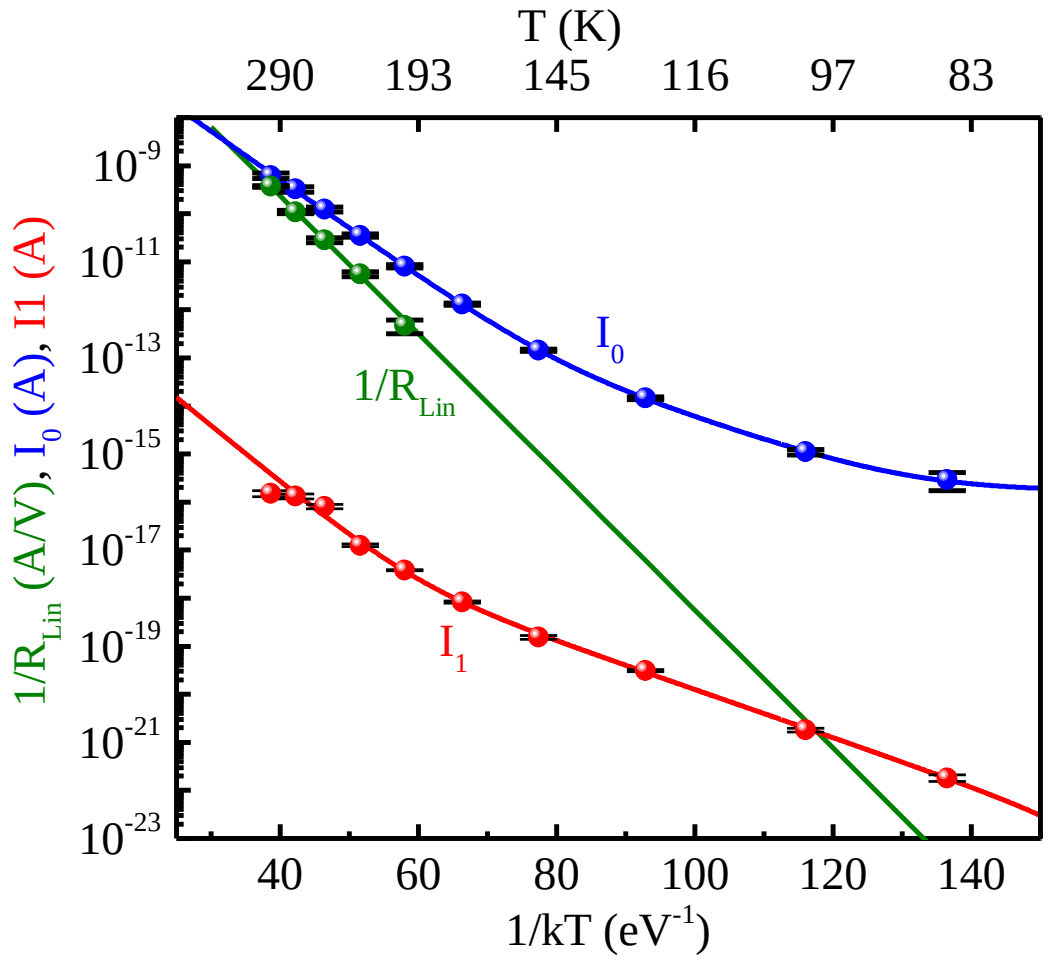

Fig 7: Temperature dependence of slope of the linear fit ($1/R_{Lin}$) and pre-factors of low ($I_0$) and high ($I_1$) field exponential fits. Activation energy of the linear ($E_{aL}$), the low ($E_{a0,1}$ , $E_{a0,2}$), and high field ($E_{a1,1}$, $E_{a1,2}$) behaviors are calculated from the slope of the linear fits of the plots.

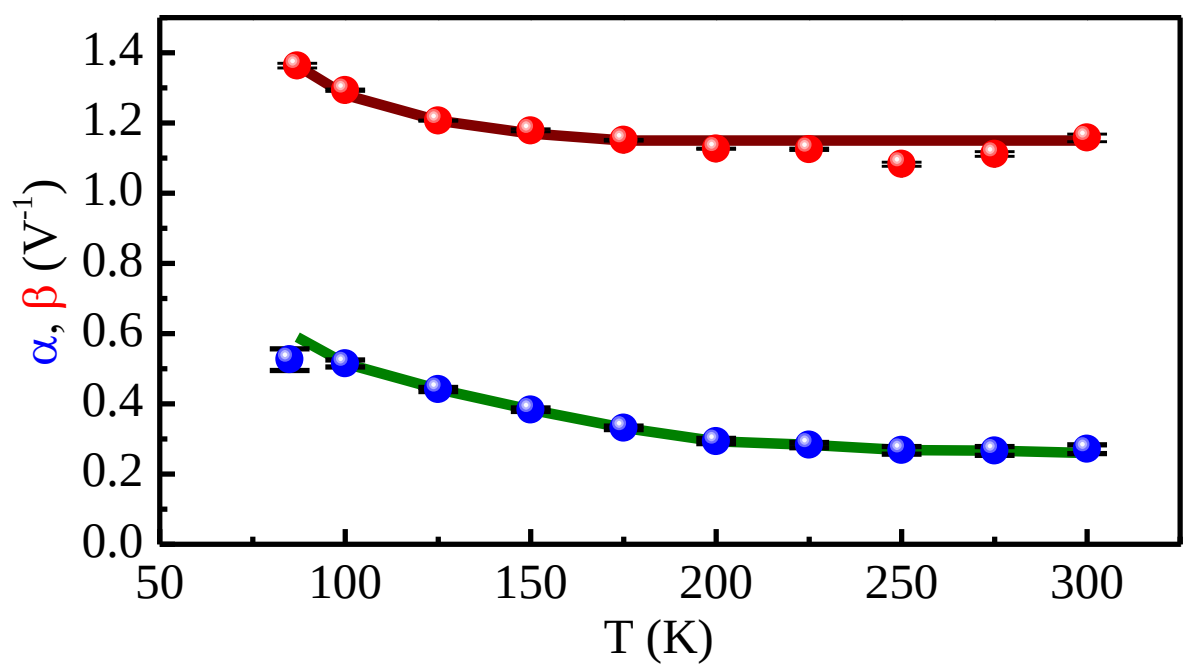

Fig 8: Exponents of low (α) and high (β) field exponential terms plotted against temperature. The solid lines show the α and β values corresponding to the fitted activation energy values (solid lines in Fig 7), stabilizing at constant values as the temperature is increased.

where $E_{aL}$ is the activation energy associated with the linear I-V behavior. We observe $E_{aL}$ to be a constant value of 331 ± 5 meV in for T < 300 K, comparable to the room temperature metastable a-GST activation energy reported by our group as well as others[21–24] (Fig 7). Current at low fields rapidly declines at lower temperatures, going below the resolution limit of our setup for T ≤ 150 K. The temperature dependent behaviors of both the low-field ($I_0$) and high-field ($I_1$) exponential prefixes (eq. 2) can be expressed as sum of two exponential terms:

$$I_0 = I_{0,1} e^{-\frac{E_{a0,1}}{kT}} + I_{0,2} e^{-\frac{E_{a0,2}}{kT}}, \quad (5)$$

$$I_1 = I_{1,1} e^{-\frac{E_{a1,1}}{kT}} + I_{1,2} e^{-\frac{E_{a1,2}}{kT}}, \quad (6)$$

This suggests that there are two independent mechanisms giving rise to the low-field and high field exponential response, one dominating at higher temperatures and the other at lower temperatures. The activation energies associated with the low field ($E_{a0,1}$ = 233 ± 2 meV, $E_{a0,2}$ = 109 ± 5) and high field ($E_{a1,1}$ = 253 ± 49 meV, $E_{a1,2}$ = 107 ± 11) behaviors have very similar values and therefore represented by fixed values ($E_{a1}$, $E_{a2}$). The exponents α and β of the voltage dependent exponential terms decrease and then stabilize with increasing temperature (Fig 8). The current-voltage relationship can be expressed as follows in the whole temperature range:

,

$$I = \frac{V}{R_0} e^{-\frac{E_{aL}}{kT}} + \left(I_{0,1} e^{-\frac{E_{a1}}{kT}} + I_{0,2} e^{-\frac{E_{a2}}{kT}}\right)\left(e^{\alpha V} - 1\right) + \left(I_{1,1} e^{-\frac{E_{a1}}{kT}} + I_{1,2} e^{-\frac{E_{a2}}{kT}}\right) e^{\beta V} + I_{offset}, \quad (7)$$

We observe a change in the low-field behavior under light, while the high-field behavior does not show a significant change (Fig 11). The effect of light is more significant at lower temperatures as there are very few thermally activated carriers.

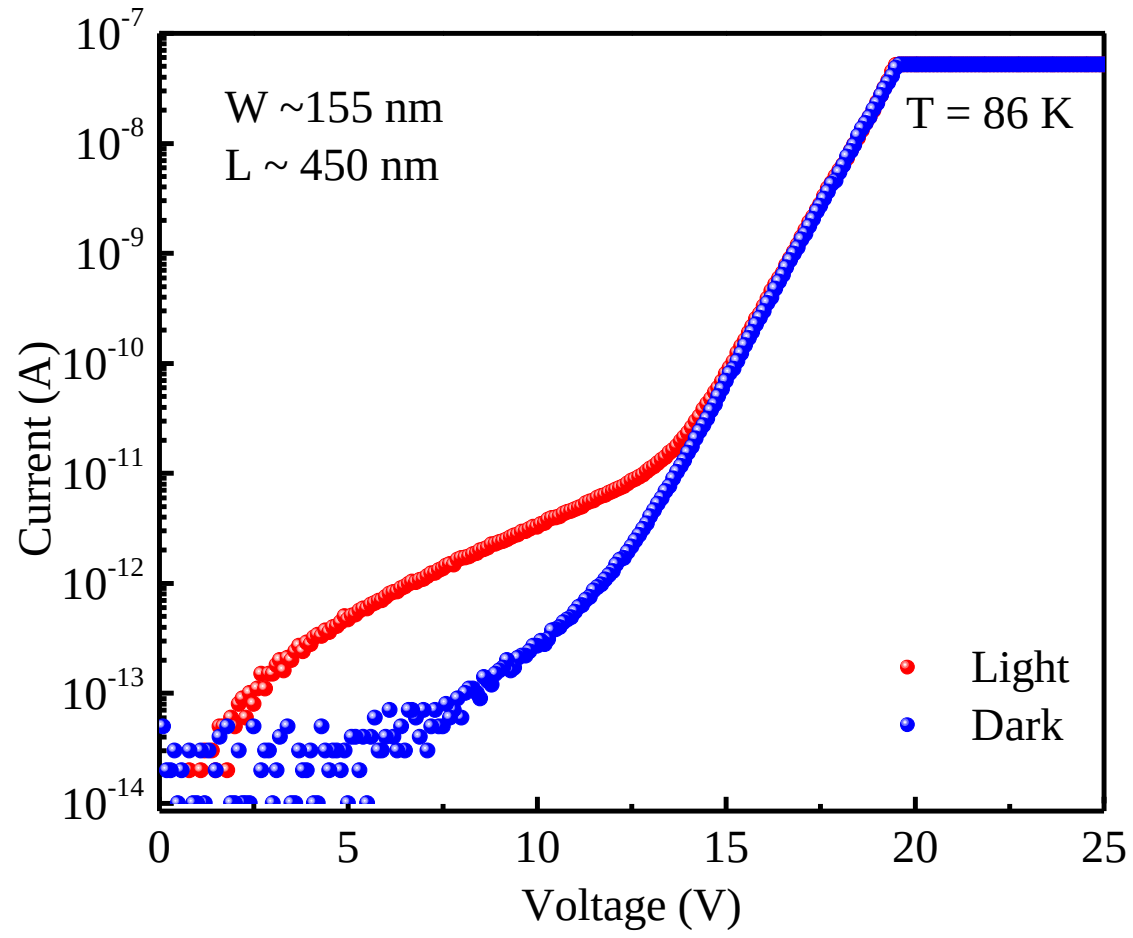

Fig 9: High-voltage sweeps (0 to 25 V) under light and dark at 86 K after the cell was stabilized with several high-voltage sweeps. Compliance was set to 50 nA.

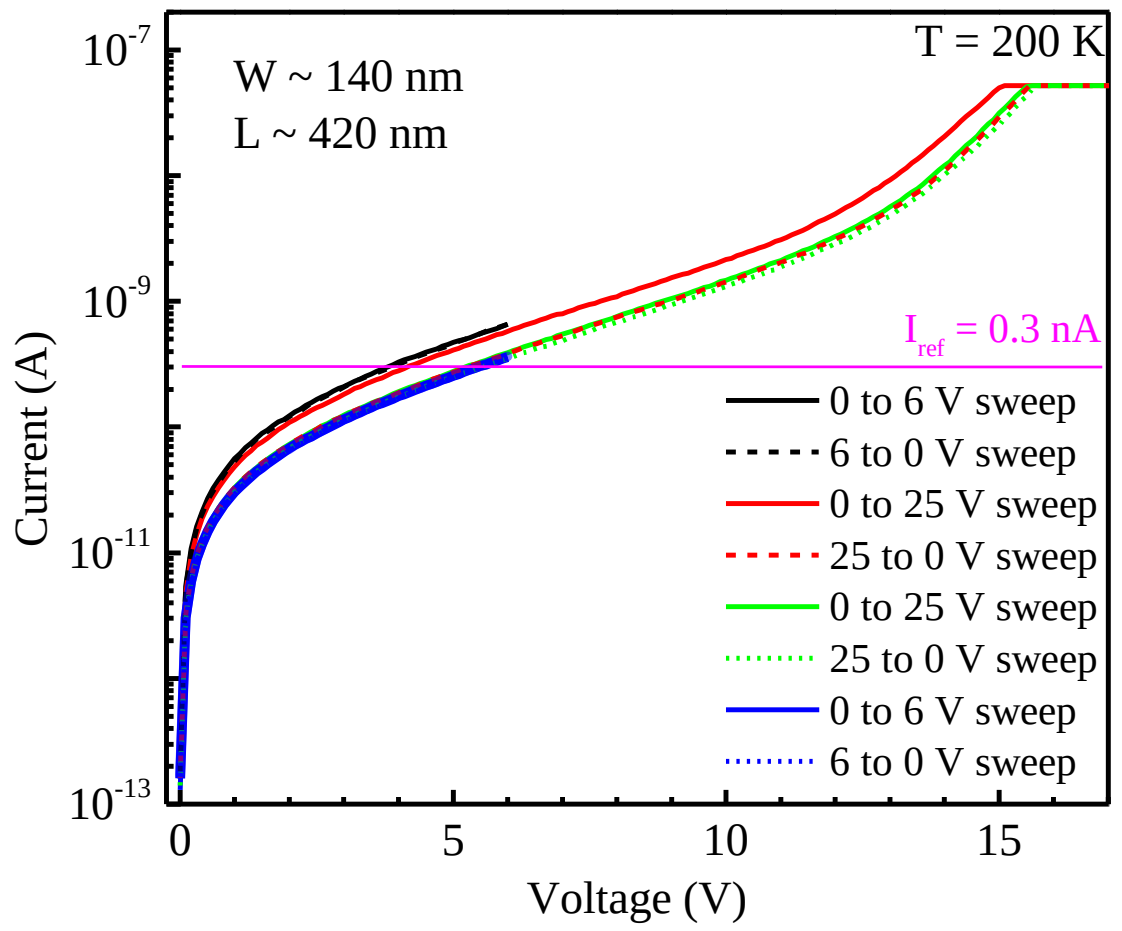

Fig 10: 0-6 V sweeps before (black) and after (blue) the 0 to 25 V sweeps (red and green).

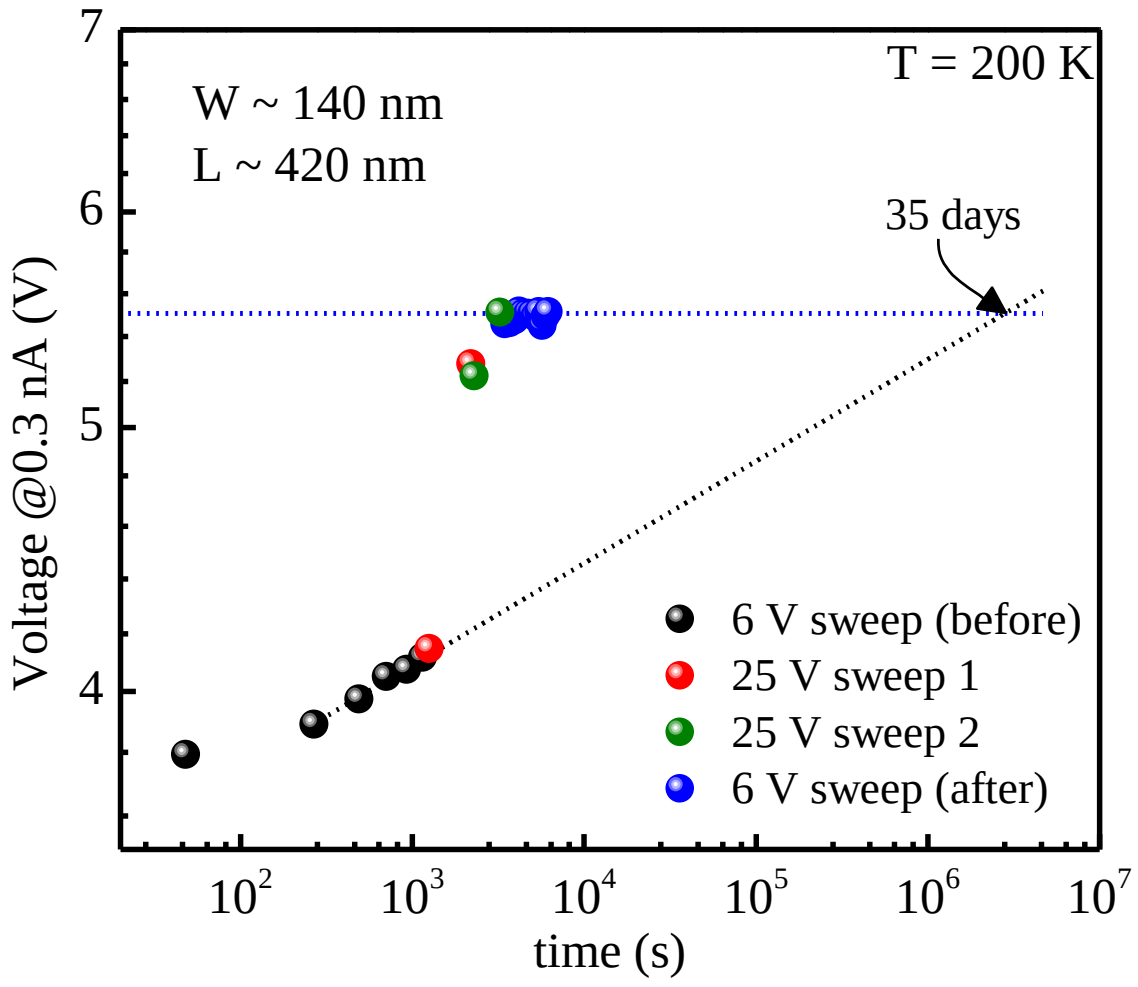

Fig 11: Bi-logarithmic plot of time versus voltage required for current to reach 0.3 nA. Required voltage increases following a power law behavior (black) before the application of 25 V sweep (red), after which the voltage become stable (blue).

In order to characterize the stability of the cells after the high-voltage sweeps at low temperatures, we performed voltage sweeps up to 6 V with $I_{comp}$ = 1 nA before and after the high-voltage sweeps (Fig 10). The 0 to 6 V sweeps do not impact the state of the cells and enable us to determine the voltage required to reach a reference current level ($I_{ref}$). We have observed that two 0 to 25 V hysteresis sweeps can stabilize the cells to a level that is expected to be reached in 35 days at 200 K (Fig 11). In other words, application of high-voltages for a few minutes can accelerate drift to reach a stable value that is normally expected in time-scale of months. Cells stabilized with a positive high-voltage sweep do not change with the application of a reverse polarity high-voltage sweep. Hence, we think that the changes are not at the contact regions but in the bulk of the devices.

The cells typically do not show significant hysteresis or acceleration in drift close to room temperature with $I_{comp}$ = 50 nA. We also do not observe full stability even if the $I_{comp}$ is increased to 1 µA at 300 K, although we do observe a hysteresis and slowing of resistance drift (Fig. 3).

In conclusion, presence of resistance drift at cryogenic temperature with drift coefficients similar to room temperature in the 200 K to 300K range and acceleration and stoppage of drift with application of high-voltage stress with low-current densities suggest that resistance drift is predominantly an electronic process for T < 300 K. High electric-fields may increase de-trapping of the excess charges that are left in the material during the quenching process. The change in the electrostatic potential profile inside the cells may be giving rise to increased cell resistance. Since the cells naturally drift to their high-resistance state without application of an external field, the resistance drift is expected to be a charge relaxation process that can be accelerated with high-fields. In our experiments we observe electric fields in the order of 70% of the room temperature breakdown fields to give rise to significant acceleration of drift. These experimental results suggest that it may be possible to minimize resistance drift by waveform engineering to reduce the number of trapped charges after reset operations, and by device engineering to reduce the effect of the trapped charges on the resistance of the devices by increasing device capacitance. Changes in the device capacitance may also be the reason for the differences observed in resistance drift in encapsulated and freestanding phase change nanowires[25].

**Acknowledgement**

Raihan Khan and Hasan Talukder performed the experiments, analysis, and writing of the manuscript supported by US NSF through grant # ECCS 1711626. The devices were fabricated at IBM T.J. Watson Research Center under a joint study agreement, by Faruk Dirisaglik supported by US DOE Office of Basic Energy Sciences (BES) and Turkish Educational Ministry. Ali Gokirmak and Helena Silva contributed to the design of experiments, analysis and writing of the manuscript. The authors are grateful to contributions of Adam Cywar of University of Connecticut, supported through US NSF Graduate Research Fellowship, and Chung Lam of IBM Research, for device fabrication. The authors would also like to thank Sadid Muneer and Nafisa Noor of University of Connecticut for valuable discussions.

## References

[1] Intel Newsroom, (2018).

[2] H.P. Wong, S. Raoux, S. Kim, J. Liang, J.P. Reifenberg, B. Rajendran, M. Asheghi, and K.E. Goodson, Proc. IEEE **98**, 2201 (2010).

[3] M. Wuttig and N. Yamada, Nat. Mater. **6**, 824 (2007).

[4] G. Bakan, N. Khan, A. Cywar, K. Cil, M. Akbulut, A. Gokirmak, and H. Silva, J. Mater. Res. **26**, 1061 (2011).

[5] G. Bakan, N. Khan, H. Silva, and A. Gokirmak, Sci. Rep. **3**, (2013).

[6] G.W. Burr, P. Tchoulfian, T. Topuria, C. Nyffeler, K. Virwani, A. Padilla, R.M. Shelby, M. Eskandari, B. Jackson, and B.S. Lee, in *J. Appl. Phys.* (2012).

[7] G. Eising, T. Van Damme, and B.J. Kooi, Cryst. Growth Des. **14**, 3392 (2014).

[8] A. Faraclas, N. Williams, A. Gokirmak, and H. Silva, IEEE Electron Device Lett. **32**, 1737 (2011).

[9] W. Kim, M. Brightsky, T. Masuda, N. Sosa, S. Kim, R. Bruce, F. Carta, G. Fraczak, H.Y. Cheng, A. Ray, Y. Zhu, H.L. Lung, K. Suu, and C. Lam, in *Tech. Dig. - Int. Electron Devices Meet. IEDM* (2017), pp. 4.2.1-4.2.4.

[10] A. Pirovano, A.L. Lacaita, F. Pellizzer, S.A. Kostylev, A. Benvenuti, and R. Bez, IEEE Trans. Electron Devices **51**, 714 (2004).

[11] M. Nardone, M. Simon, I. V. Karpov, and V.G. Karpov, J. Appl. Phys. (2012).

[12] F. Dirisaglik, High-Temperature Electrical Characterization of $Ge_2Sb_2Te_5$ Phase Change Memory Devices, University of Connecticut, 2014.

[13] R.S. Khan, F. Dirisaglik, A. Gokirmak, and H. Silva, ArXiv:1912.04480 (2019).

[14] D. Ielmini, D. Sharma, S. Lavizzari, and A.L. Lacaita, in *IEEE Int. Reliab. Phys. Symp. Proc.* (2008), pp. 597–603.

[15] M. Boniardi and D. Ielmini, Appl. Phys. Lett. **98**, 1 (2011).

[16] A. Sebastian, D. Krebs, M. Le Gallo, H. Pozidis, and E. Eleftheriou, in *IEEE Int. Reliab. Phys. Symp. Proc.* (2015).

[17] S. Kim, B. Lee, M. Asheghi, F. Hurkx, J.P. Reifenberg, K.E. Goodson, and H.S.P. Wong, IEEE Trans. Electron Devices (2011).

[18] F. Dirisaglik, G. Bakan, Z. Jurado, S. Muneer, M. Akbulut, J. Rarey, L. Sullivan, M. Wennberg, A. King, L. Zhang, R. Nowak, C. Lam, H. Silva, and A. Gokirmak, Nanoscale **7**, 16625 (2015).

[19] D. Ielmini, S. Lavizzari, D. Sharma, and A.L. Lacaita, in *Tech. Dig. - Int. Electron Devices Meet. IEDM* (2007), pp. 939–942.

[20] D.S. Sanditov and S.B. Munkueva, Glas. Phys. Chem. **42**, 135 (2016).
[21] D. Ielmini and Y. Zhang, Electron Devices Meet. 2006. IEDM’06. (2006).
[22] D. Ielmini and Y. Zhang, J. Appl. Phys. **102**, 54517 (2007).
[23] M. Wimmer, M. Kaes, C. Dellen, and M. Salinga, Front. Phys. **2**, 1 (2014).
[24] S. Muneer, J. Scoggin, F. Dirisaglik, L. Adnane, A. Cywar, G. Bakan, K. Cil, C. Lam, H. Silva, and A. Gokirmak, AIP Adv. **8**, (2018).
[25] M. Mitra, Y. Jung, D.S. Gianola, and R. Agarwal, Appl. Phys. Lett. (2010).